\documentstyle[aps,prl,preprint,floats,epsfig]{revtex}

\begin{document}

\preprint{\tighten\vbox{\hbox{\hfil CLNS 00/1682}
                        \hbox{\hfil CLEO 00-14}
}}

\title{ First Observation of the Decays 
$B^0 \rightarrow D^{*-} p \bar{p} \pi^+$ and 
$B^0 \rightarrow D^{*-} p \bar{n} $ }

\author{CLEO Collaboration}
\date{\today}

\maketitle
\tighten

\begin{abstract} 
We report the first observation of exclusive decays of the 
type $B \rightarrow D^*N\bar{N}X$, where $N$ 
is a nucleon. Using a sample of 9.7 $\times$ $10^{6}$ 
$B\bar{B}$ pairs collected with the CLEO detector 
operating at the Cornell Electron Storage Ring, 
we measure the branching fractions ${\cal B}$
({$B^0$$\rightarrow$ $D^{*-}$ $p$ ${\bar{p}}$ ${\pi}^+$}) = 
(${6.5}^{+1.3}_{-1.2}$ $\pm$ 1.0) $\times$ $10^{-4}$ and 
${\cal B}$({
$B^0$$\rightarrow$ $D^{*-}$ $p$ ${\bar{n}}$
}) = (${14.5}^{+3.4}_{-3.0}$ $\pm$ 2.7) 
$\times$ $10^{-4}$. Antineutrons are identified by their 
annihilation in the CsI electromagnetic 
calorimeter.

\end{abstract}

\newpage

{
\renewcommand{\thefootnote}{\fnsymbol{footnote}}

\begin{center}
S.~Anderson,$^{1}$ V.~V.~Frolov,$^{1}$ Y.~Kubota,$^{1}$
S.~J.~Lee,$^{1}$ R.~Mahapatra,$^{1}$ J.~J.~O'Neill,$^{1}$
R.~Poling,$^{1}$ T.~Riehle,$^{1}$ A.~Smith,$^{1}$
C.~J.~Stepaniak,$^{1}$ J.~Urheim,$^{1}$
% ...
\end{center}
 
\small
\begin{center}
$^{1}${University of Minnesota, Minneapolis, Minnesota 55455}\\
% ...
\end{center}

\setcounter{footnote}{0}
}
\newpage


A unique feature of the $B$ meson system is 
that the large mass of the b-quark allows for many 
of the weak decays of the $B$ meson to include 
the creation of a baryon-antibaryon pair. 
In the simplest picture, baryons
are expected to be produced in decays 
of the type $B\to{\overline\Lambda_c}pX$, 
and it has been only decays of this type which have been 
exclusively reconstructed to date \cite{jjo}. 
However, one can combine the recently measured value 
${\cal B}(\Lambda_c^+\to pK^-\pi^+) =
(5.0\pm0.5\pm1.2) \%$ \cite{russ-paper} with estimates
of the product branching fraction
${\cal B}(B\to\Lambda_c X)\times
{\cal B}(\Lambda_c\to pK^-\pi^+)$ of 
$(1.81\pm0.22\pm0.24)\times 10^{-3}$\cite{zoeller}
to determine that $B\to{\overline\Lambda_c} N X $ modes, where 
$N$ is a proton or a neutron, account for only 
about half of the total $B\to Baryons$ rate. 
Dunietz \cite{dunietz} has suggested that modes 
of the type {$B \rightarrow DN{\bar{N}} X $},
in which $D$ represents any 
charmed meson, are 
likely to be sizeable. $B \rightarrow DN{\bar{N}} X $ 
final states can arise from either the hadronization 
of the $W$ boson into a baryon-antibaryon pair, or 
the production of a highly excited 
charmed baryon which decays strongly into 
a baryon plus a charmed meson. 
CLEO has previously reported an inclusive 
upper limit for ${\cal B}$($B \rightarrow DN{\bar{N}} X $) 
of $<$ 4.8 $\%$
at 90$\%$ confidence level 
\cite{cleo92}. 
We report the first observation of decays of this type, and present
measurements of the branching fractions 
${\cal B}$({$B^0$$\rightarrow$ $D^{*-}$ $p$ ${\bar{p}}$ ${\pi}^+$}) 
and ${\cal B}$({$B^0$$\rightarrow$ $D^{*-}$ $p$ ${\bar{n}}$}). 
The charge conjugate process is implied in 
the reconstruction of 
$B^0$$\rightarrow$ $D^{*-}$ $p$ ${\bar{p}}$ ${\pi}^+$. 
However, in the reconstruction of 
$B^0$$\rightarrow$ $D^{*-}$ $p$ ${\bar{n}}$ 
only the mode with the antineutron is used in our 
measurement because neutrons do not have a
distinctive annihilation signature in the CLEO detector.


The data were taken with the CLEO II detector 
\cite{detector} at the Cornell Electron Storage Ring (CESR). 
The sample we use corresponds to an integrated luminosity 
of 9.1 fb$^{-1}$ from data taken on the $\Upsilon(4S)$ 
resonance, corresponding to 
$9.7 \times 10^6 B\bar{B}$ pairs, and 4.5 fb$^{-1}$ in the 
continuum at energies just below the $\Upsilon(4S)$. 
We assume
that $50\%$ of the $B\bar{B}$ pairs consist of $B^0\bar{B^0}$, 
and that there are equal numbers of $B^0$ and $\bar{B^0}$ mesons. 
Charged particle trajectories are measured in a cylindrical 
drift chamber operating in a 1.5 T magnetic field. 
Photons and antineutrons are detected using an 
calorimeter consisting of 7800 CsI 
crystals with excellent resolution in position and 
electromagnetic shower energy. 
Simulated events were generated with a GEANT-based
Monte Carlo program \cite{geant}.
Sixty percent of the data were taken in the CLEO II.V
configuration \cite{thill}.

Charged particle identification is accomplished 
by combining the specific ionization (dE/dx) measurements 
from the drift chamber with time-of-flight 
(TOF) scintillation counter measurements. 
We reconstruct the decay mode 
$D^{*-} \rightarrow \bar{D}^0 \pi^-$, with 
$\bar{D}^0$$\rightarrow$$K^+$${\pi}^-$, 
$\bar{D}^0$$\rightarrow$$K^+$${\pi}^-$${\pi}^0$, and 
$\bar{D}^0$$\rightarrow$$K^+$${\pi}^-$${\pi}^+$${\pi}^-$. 
Pairs of calorimeter showers with 
photon-like lateral shower shapes and invariant 
mass within 2.5 standard deviations of $M({\pi^0})$ 
are considered as $\pi^0$ candidates. We select $D^{*-}$
candidates using 95\%
efficient cuts around the 
central values of the $M_{D^0}$ 
and $(M_{D^*} - M_{D^0})$ distributions. 
In the few cases where there is more than one {$D^*$}
candidate in an event, the candidate which is closest 
to $M_{D^0}$ = 1.8646 GeV and 
$(M_{D^*} - M_{D^0})$ = 0.1454 GeV \cite{pdg98} is chosen.


To reconstruct $B^0$ candidates for {
$B^0$ $\rightarrow$ $D^{*-}$ $p$ ${\bar{p}}$ ${\pi}^+,$} we 
calculate the beam constrained mass, $M(B) \equiv 
\sqrt{ {E_{\rm beam}}^2 - p(B)^2}$, where 
$E_{\rm beam}$ is the beam energy, and 
$p(B)$ is the $B^0$ candidate three-momentum magnitude. 
We also use the energy difference 
between the beam energy and the energy of the 
reconstructed $B$ candidate: 
$\Delta E \equiv E_{\rm beam} - E(B)$. 
Using a Monte Carlo simulation program, we find the 
detector resolution for the $\Delta E$ distribution 
for each $D^0$ mode, and require that the measured 
$\Delta E$ is within 3 standard deviations of zero.

\begin{figure}[ht]
   \centering \leavevmode
        \epsfysize=5cm
   \epsfbox{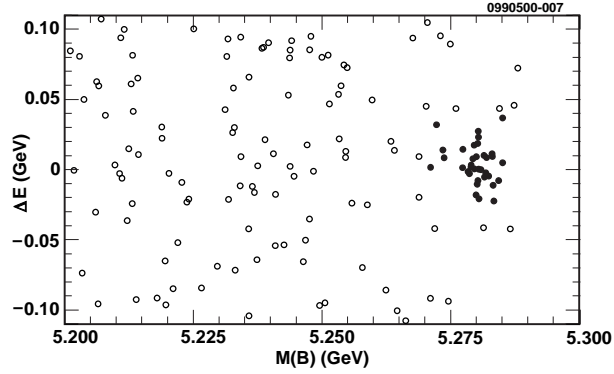}
\vskip 20pt
   \caption{  $\Delta E$ vs $M(B)$  distribution for 
$B^0$$\rightarrow$ $D^{*-}$ $p$ ${\bar{p}}$ ${\pi}^+$. The solid 
circles indicate events in the signal region.  } 
\label{dele}
\end{figure}
\begin{figure}[ht]
   \centering \leavevmode
        \epsfysize=8cm
   \epsfbox{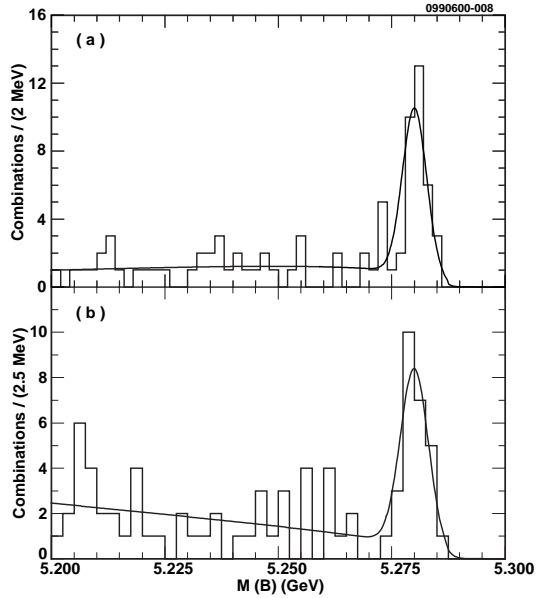}
\vskip 20pt
   \caption{ $M(B)$ distribution for (a) 
$B^0$$\rightarrow$ $D^{*-}$ $p$ ${\bar{p}}$ ${\pi}^+$ 
and (b) {$B^0$$\rightarrow$ $D^{*-}$ $p$ ${\bar{n}}$}.
Each plot is fitted to the sum of a background function and a
fixed-width Gaussian signal shape.  } 
\label{bpppidsdat}
\end{figure}

The $M(B)$ resolution is dominated by the 
beam energy spread and is consistent 
with the Monte Carlo simulation 
for $B^0$ $\rightarrow$ $D^{*-}$ $p$ ${\bar{p}}$ ${\pi}^+$, 
which predicts it to be Gaussian with a width of $\sigma$ = 2.7 MeV. 
The $M(B)$ distribution is fitted to a Gaussian function with 
the predicted width, and a polynomial background function with 
suppression at the $E_{\rm beam}$ threshold. 
The fitted signal yield is ${32.3}^{+6.3}_{-6.0}$ events, 
where the errors are statistical only. 
If $\sigma$ is allowed to float, a value of $\sigma = 2.1 \pm 0.4$ MeV
is obtained, consistent with the Monte Carlo prediction.
Figure \ref{dele} shows  $\Delta E$ vs $M(B)$  for 
$B^0$ $\rightarrow$ $D^{*-}$ $p$ ${\bar{p}}$ ${\pi}^+$, 
and Figure \ref{bpppidsdat}(a) shows $M(B)$ for 
$B^0$ $\rightarrow$ $D^{*-}$ $p$ ${\bar{p}}$ ${\pi}^+$. 

To find the decay 
$B^0$$\rightarrow$ $D^{*-}$ $p$ ${\bar{n}}$ requires the detection of an
antineutron. This is accomplished by identifying annihilation 
showers in the CsI calorimeter.
Antiprotons and antineutrons annihilate with nucleons 
in the calorimeter. 
These annihilations result in 
showers with different characteristics than those from 
photons or charged particles. 
We use antiproton annihilation showers 
to define the antineutron selection criteria 
since we are unable to isolate a sample of 
antineutrons in data. Our antiproton sample consists of 
$1.6 \times 10^5$ $\bar{p}$'s from reconstructed 
$\bar{\Lambda} \rightarrow \bar{p} \pi^+$ decays, 
in which the daughter antiprotons are selected 
by dE/dx and TOF response. 
The isolation of this sample is 
independent of calorimeter response and therefore 
allows us to evaluate our shower-based selection 
criteria. We can expect these same criteria to be 
similarly effective for antineutrons. 

\begin{figure}[ht]
   \centering \leavevmode
        \epsfysize=4.5cm
   \epsfbox{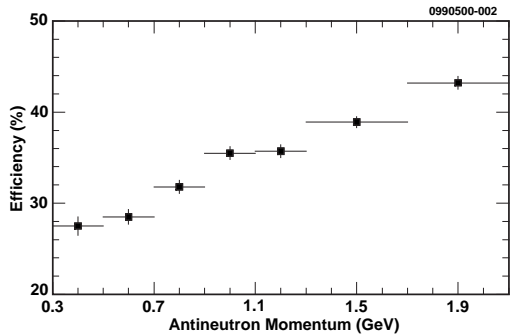}
\vskip 20pt
   \caption{ Antineutron 
reconstruction efficiency
derived from simulated
$B^0\rightarrow D^{*-}p\bar{n}$ events.
\label{anti}} 
\end{figure}

We find that antinucleons typically deposit a substantial 
amount of energy in a laterally broad shower, 
which has energy $E_{\rm main}$, and a lesser amount of 
energy in adjacent satellite showers, whose energy 
is added to $E_{\rm main}$ to define $E_{\rm group}$. 
The selection of showers with $E_{\rm main}$ $>$ 500 MeV 
and $E_{\rm group}$ $>$ 800 MeV is useful in 
suppressing backgrounds, while retaining many 
antinucleon annihilation showers. 
The main shower must have polar angle $\theta$ 
with respect to the incoming positron direction 
of $45^{\circ} < \theta < 135^{\circ}$ to ensure that it
is located in that part of the calorimeter that has the best
resolution, must have a lateral shape broader than typical 
photons, and must not match the projection of 
any charged track trajectory. 
Baryon number conservation serves as an 
added background suppressant: once we require that 
there be a proton, the likelihood of 
an annihilation-like shower in the event to be 
an antineutron increases significantly. 
The reconstruction efficiency 
as a function of momentum 
for antineutrons in a Monte Carlo sample of 
$B^0$$\rightarrow$ $D^{*-}$ $p$ ${\bar{n}}$ events 
in which the $B^0$ 
selection criteria have been applied is shown in Figure \ref{anti}.

The measured shower 
energy for an antineutron, $E_{\rm group}$,  
does not give an accurate measurement of the 
total energy of the antineutron. 
We therefore assign the energy
of the antineutron candidate using 
$E(\bar{n}) = E_{\rm beam} - E(D^{*-}) - E(p)$. 
We then use this energy, together with the
position of the antineutron shower, to calculate the
momentum of the antineutron candidate.
This momentum is added to the momenta of the $D^{*-}$ and $p$
candidates to give the momentum of the $B^0$ meson, $p(B)$.
The $B^0$ candidate mass is then calculated with 
$M(B)$ = $\sqrt{E_{\rm beam}^2-p(B)^2}$.


The mass resolution of the reconstructed $B^0$ meson is estimated 
from a Monte Carlo simulation to be Gaussian with a width 
$\sigma$ = 3.1 MeV, demonstrating 
that the lack of a direct measurement of the antineutron energy
does not seriously degrade the $M(B)$ resolution relative 
to $B^0$ $\rightarrow$ $D^{*-}$ $p$ ${\bar{p}}$ ${\pi}^+$. However, 
for this mode we cannot use the requirement that $\Delta E$ is
consistent with zero, as we do not have two independent measures
of the energy of the $B^0$. 


In reconstructing {$B^0$$\rightarrow$ $D^{*-}$ $p$ ${\bar{n}}$} 
we could also be reconstructing 
{$B^0$ $\rightarrow$ $D^{*-}$ ${D^{+}_{s}}$} 
with {${D^{+}_{s}}$ $\rightarrow$ $p$ ${\bar{n}}$}. 
This latter decay has not been observed but could 
occur via $c\bar{s}$ annihilation. 
We might also be including events that are 
{$B^0$ $\rightarrow$ $D^{*-}$ ${D^{*+}_{s}}$} 
with {$D^{*+}_{s} \rightarrow D^{+}_{s} \gamma$}
or {$D^{*+}_{s} \rightarrow D^{+}_{s} \pi^0$} and 
{${D^{+}_{s}}$ $\rightarrow$ $p$ ${\bar{n}}$}. 
These events will populate the $M(B)$ signal region but with a broader 
signal peak due to the missing soft photon or $\pi^0$. We eliminate 
these two types of decay by rejecting events with 
1.91 $<$ $M(p  \bar{n})\ $(GeV) $<$ 2.04 GeV, for a 
loss of only 9$\%$ in the relative reconstruction efficiency. 


The final data $M(B)$ distribution, shown in 
Fig. \ref{bpppidsdat}(b), is fit with a 
Monte Carlo predicted Gaussian width of 
$\sigma$ = 3.1 MeV and gives a signal yield 
of ${24.0}^{+5.6}_{-5.0}$ events. 
If $\sigma$ is allowed to float, 
a value of 2.6 $\pm$ 0.4 MeV is found, 
consistent with the Monte Carlo expectation.


We use a Monte Carlo simulation to calculate detection 
efficiencies to be 0.52$\%$ 
for $B^0$$\rightarrow$ $D^{*-}$ $p$ ${\bar{p}}$ ${\pi}^+$, 
and 0.34$\%$ for 
$B^0$$\rightarrow$ $D^{*-}$ $p$ ${\bar{n}}$, where these numbers
include all the relevant branching fractions of the daughters. 
The Monte Carlo generation of the decays assume no 
resonant sub-structure. 
No evidence of sub-structure has been found in the signal candidates.
However, we will allow for possibility
of sub-structure, which would alter the efficiency, 
in our estimation of the systematic uncertainties. 

\begin{figure}[ht]
   \centering \leavevmode
        \epsfysize=8cm
   \epsfbox{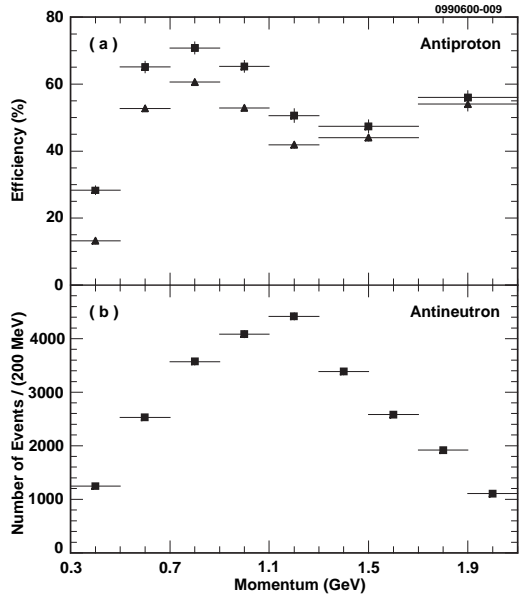}
\vskip 20pt
   \caption{ (a) Antiproton 
reconstruction efficiency. Squares represent 
Monte Carlo efficiency and triangles 
that of the data. (b) Monte Carlo antineutron momentum 
spectrum in simulated $B^0 \rightarrow D^{*-}p\bar{n}$ decays.
\label{prot}} 
\end{figure}

Whereas the Monte Carlo simulation has been well 
tested for charged particles and photons, 
this is the first measurement that has explicitly 
needed the efficiency for antinucleon annihilations in the 
calorimeter. We find a discrepancy 
for antiproton annihilation showers 
between the Monte Carlo and data.
The 
reconstruction efficiency for Monte Carlo and 
data for antiprotons, as determined from the 
aforementioned $\bar{\Lambda}$ sample, 
is shown in Fig. \ref{prot}(a). 
We find that the antiproton reconstruction efficiency 
is overestimated by the GEANT simulation prediction. 
We assume that the 
Monte Carlo simulation for the antineutron efficiency 
is similarly overestimated and therefore must be corrected. 
Weighting the efficiency correction for antiprotons 
by the expected antineutron momentum spectrum results 
in an antineutron selection 
efficiency which is 21$\%$ lower than the 
GEANT Monte Carlo simulation. In Fig. \ref{prot}(b) 
we show the antineutron momentum spectrum generated 
in Monte Carlo for $B\to D^{*-}p\bar{n}$ decays.

We search for, but do not find,  
resonant substructure contributions to 
$B^0\rightarrow D^{*-} p \bar{p} \pi^+$
and $B^0\rightarrow D^{*-} p \bar{n}$ decays. Examples
of possible substructure arise
from a heavy charmed baryon decaying 
strongly to ($\bar{p}$ $D^{*-}$) for 
$B^0$$\rightarrow$ $D^{*-}$ $p$ ${\bar{p}}$ ${\pi}^+$ 
and ($\bar{n}$ $D^{*-}$) for 
$B^0$$\rightarrow$ $D^{*-}$ $p$ ${\bar{n}}$, 
and a resonance of the virtual $W$ 
decaying to $p \bar{p} \pi^+$. 
We also study the 
effect on the Monte Carlo reconstruction efficiency 
of a two-body decay into $D^{*-}X$,  
and find no  evidence from the kinematic 
distribution of the daughter particles for 
decays of this type. We also find insignificant background 
from $B$ decays with a charmed baryon final state 
or from other $B \rightarrow D^{*-} X$ modes.


We have considered many sources of 
systematic uncertainty of these
branching fraction measurements. 
Systematic uncertainties shared by 
both modes are statistical uncertainty of 
$D^0$ branching fractions (0.6$\%$) and 
$D^{*-}$ branching fraction (1.4$\%$), 
$D^{*-}$ reconstruction due to kinematic 
fitting and particle identification (5.0$\%$), 
and Monte Carlo statistics (5.0$\%$). 
Systematic uncertainties which differ for the 
two modes, and which we quote in parenthesis 
for 
$B^0$$\rightarrow$ $D^{*-}$ $p$ ${\bar{p}}$ ${\pi}^+$ 
and $B^0$$\rightarrow$ $D^{*-}$ $p$ ${\bar{n}}$, 
respectively, are: tracking, 1$\%$ per track 
(6.6$\%$, 4.6$\%$), proton identification 
criteria (8$\%$, 4$\%$), and n-body versus 
two-body for decay kinematics (5$\%$, 3$\%$). 
No statistically significant $\Delta$ baryon contribution
to the $D^{*-}$ $p$ ${\bar{p}}$ ${\pi}^+$ yield
was found. We place a systematic uncertainty due to a 
possible $\Delta$ baryon contribution to the 
$B^0$$\rightarrow$ $D^{*-}$ $p$ ${\bar{p}}$ ${\pi}^+$ 
signal. Also, we assign a 5\% uncertainty on the yield
of 
$B^0$$\rightarrow$ $D^{*-}$ $p$ ${\bar{n}}$ 
due to the possibility of $\Delta$ baryons with
a missing $\pi$ distorting the background shape in this mode. 
The systematic uncertainty for 
antineutron identification is estimated to be 15$\%$.
The quadrature sum of all systematic uncertainties 
is 15$\%$ for 
$B^0$$\rightarrow$ $D^{*-}$ $p$ ${\bar{p}}$ ${\pi}^+$, 
and 19$\%$ for 
$B^0$$\rightarrow$ $D^{*-}$ $p$ ${\bar{n}}$.


In conclusion, we have made the first observation of decay
modes of the $B^0$ of the type $B \to DN\bar{N}X$, 
which may contribute substantially to the total 
observed $B \rightarrow Baryons$ rate. 
We measure the branching fractions ${\cal B}$
({$B^0$$\rightarrow$ $D^{*-}$ $p$ ${\bar{p}}$ ${\pi}^+$}) = 
(${6.5}^{+1.3}_{-1.2}$ $\pm$ 1.0) $\times$ $10^{-4}$, and 
${\cal B}$({
$B^0$$\rightarrow$ $D^{*-}$ $p$ ${\bar{n}}$
}) = (${14.5}^{+3.4}_{-3.0}$ $\pm$ 2.7) 
$\times$ $10^{-4}$. 

\smallskip
We gratefully acknowledge the effort of the 
CESR staff in providing us with
excellent luminosity and running conditions.
This work was supported by 
the National Science Foundation,
the U.S. Department of Energy,
the Research Corporation,
the Natural Sciences and Engineering Research Council of Canada, 
the A.P. Sloan Foundation, 
the Swiss National Science Foundation, 
and the Alexander von Humboldt Stiftung. 


\end{document}